\documentclass{article}
\usepackage{spconf,amsmath,graphicx}
\usepackage{booktabs}

\newcommand{\Figure}[3]{\vspace{-0mm} \includegraphics[width=#1,clip]{#2.eps} \vspace{-3mm} \caption{#3} \vspace{-5mm} \label{fig:#2}} 

\renewcommand{\Vec}[1]{\textrm{\boldmath $#1$}} 

\newcommand{\y}{ \Vec{y} } 
\newcommand{\vecs}{ \Vec{s} } 
\newcommand{\mats}{ \Vec{S} } 
\newcommand{\n}{ \Vec{n} } 
\newcommand{\haty}{ \Vec{\hat y} } 

\newcommand{\drawfig}[4]{ 
  \begin{figure}[#1]
  \begin{center}
  \Figure{#2}{#3}{#4} 
  \end{center} 
  \end{figure}
}

\title{GENERATIVE MOMENT MATCHING NETWORK-BASED RANDOM MODULATION POST-FILTER FOR DNN-BASED SINGING VOICE SYNTHESIS AND NEURAL DOUBLE-TRACKING}

\makeatletter
\def\name#1{\gdef\@name{#1\\}}
\makeatother
\name{{\em Hiroki Tamaru$^{1}$, Yuki Saito$^{1}$, Shinnosuke Takamichi$^{1}$, Tomoki Koriyama$^{2}$, and Hiroshi Saruwatari$^{1}$}}

\address{$^1$ Graduate School of Information Science and Technology, The University of Tokyo, Japan. \\
$^2$ School of Engineering, Tokyo Institute of Technology, Japan. \\
}

\begin{document}
\ninept
\maketitle

\setlength{\abovedisplayskip}{3pt} 
\setlength{\belowdisplayskip}{3pt} 

\begin{abstract}

This paper proposes a generative moment matching network (GMMN)-based post-filter that provides inter-utterance pitch variation for deep neural network (DNN)-based singing voice synthesis. The natural pitch variation of a human singing voice leads to a richer musical experience and is used in double-tracking, a recording method in which two performances of the same phrase are recorded and mixed to create a richer, layered sound. However, singing voices synthesized using conventional DNN-based methods never vary because the synthesis process is deterministic and only one waveform is synthesized from one musical score. To address this problem, we use a GMMN to model the variation of the modulation spectrum of the pitch contour of natural singing voices and add a randomized inter-utterance variation to the pitch contour generated by conventional DNN-based singing voice synthesis. Experimental evaluations suggest that 1) our approach can provide perceptible inter-utterance pitch variation while preserving speech quality. We extend our approach to double-tracking, and the evaluation demonstrates that 2) GMMN-based neural double-tracking is perceptually closer to natural double-tracking than conventional signal processing-based artificial double-tracking is.

\end{abstract}


\section{Introduction} \label{sec:intro}
\vspace{-4pt}

These days, synthesized singing voices are being used for creating music. In particular, there are a variety of singing voice synthesis systems that work on the basis of unit selection synthesis (e.g., Vocaloid \cite{kenmochi07_vocaloid}), hidden Markov models (HMMs) \cite{saino2006hmm,oura2010recent}, and deep neural networks (DNNs) \cite{nishimura2016singing,blaauw2017neural}. One of the aims of these systems is to create expressive singing voices and music regardless of the users' gender or skill. Among them, DNN-based ones utilize a machine-learning-based synthesis process and have the potential to synthesize high-quality expressive voices.

However, the conventional DNN-based singing voice synthesis lacks \textit{inter-utterance variation}, as shown in Fig.~\ref{fig:eps/concept_1027}. Given one musical score, a human singer will sing it differently when asked to repeat it. The inter-utterance variation contributes to a richer musical experience. For instance, it proves that a singer is actually singing, not lip-synching. It also provides a music producer with the opportunity to choose a favorite from various recordings of the same song.
In contrast, only one voice is synthesized from one musical score in conventional DNN-based singing voice synthesis because the synthesis process is deterministic.
The lack of inter-utterance variation loses not only the rich musical experiences mentioned above, but also the ability to take advantage of \textit{double-tracking} (DT) \cite{brice2001music,womack2014beatles} (Fig.~\ref{fig:eps/DT_ADT_NDT_1024}). DT involves two or more vocal performances uttered by one human singer that are combined in a mix. It gives layeredness and richness to the resulting voices, thanks to the singer's inter-utterance variation. An alternative way is signal processing-based \textit{artificial double-tracking} (ADT). Instead of recording multiple performances, ADT modulates one voice (e.g., through a delay and chorus effect) and mixes the original and modulated voices. Since ADT does not require multiple voices with inter-utterance variations, it can be easily applied to not only human voices but also synthesized voices. However, the effects of ADT are known to result in unnatural sound artifacts, such as tonal alterations and timbre coloration \cite{izhaki2017mixing}.

\drawfig{t}{0.95\linewidth}{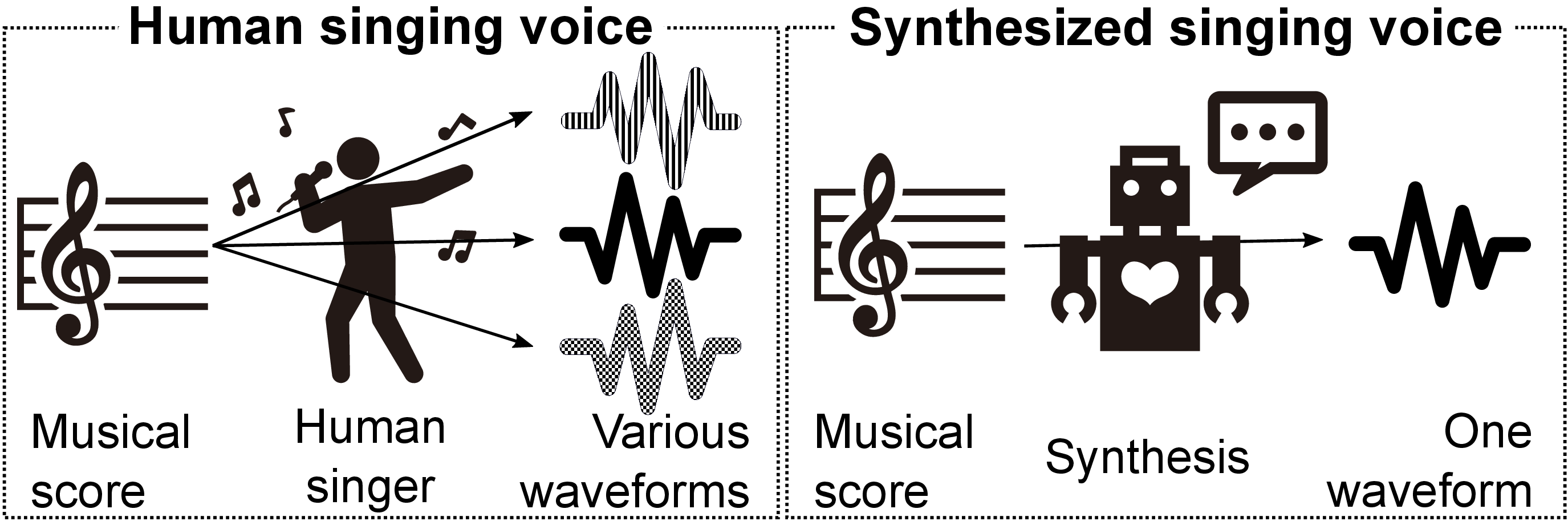} {Comparison of human and synthesized singing voices.}
\drawfig{t}{0.95\linewidth}{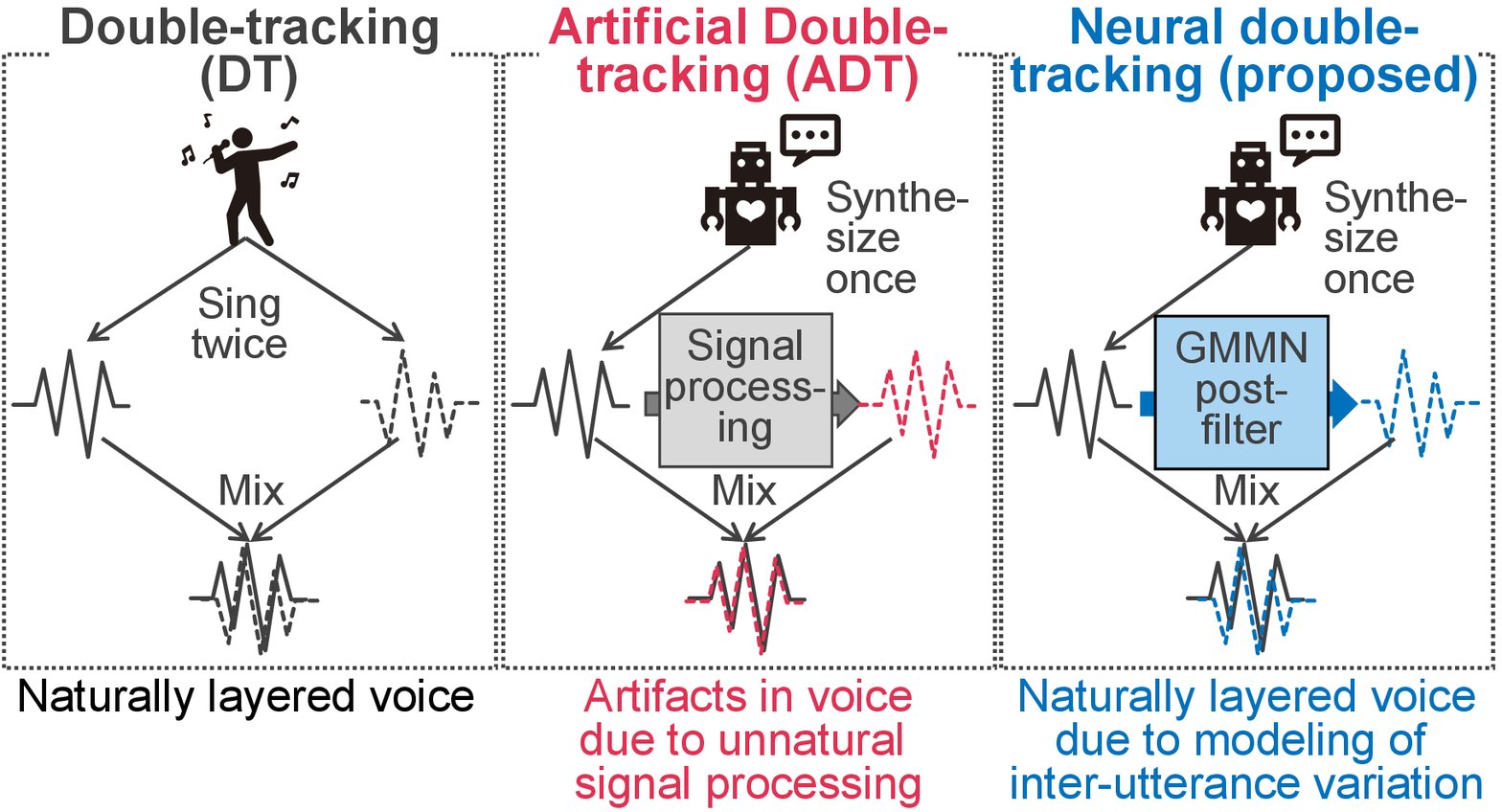} {Double-tracking (DT), artificial double-tracking (ADT), and proposed neural double-tracking. This figure shows ADT performed on synthesized voice, but it can also be performed on natural voice.}
 
To address these problems, we propose a post-filter that gives inter-utterance pitch variations to a synthesized singing voice.
Since such variations can be assumed to follow a certain complicated distribution, we use deep generative models, which are known to be capable of modeling complicated distributions.
Building on our previous work on spectrum generation in text-to-speech synthesis \cite{takamichi17moment}, we utilize a generative moment matching network (GMMN) \cite{li15momentmatchingnetwork,ren16conditionalmomentmatching} as the deep generative model because it is effective and easy to implement compared with other models. 
For example, generative adversarial networks (GANs) \cite{goodfellow14gan} involve a difficult minimax problem and variational auto-encoders \cite{kingma2013vae} are subject to the degration of the decoding quality due to over-regularization \cite{saito2018non}.
In our method, given the synthesized pitch contour and a prior noise vector, the conditional GMMN is trained to represent the distribution of natural pitch contours (i.e., the natural variation of the human singer's pitch contours). To capture the long-term pitch structure, the modulation spectrum (MS) \cite{takamichi16mspf} of the pitch contours is used, and the conditional GMMN models the natural MS variation. In the synthesis, given a prior noise vector, the GMMN randomly generates MSs that have natural inter-utterance variations, and the filtered pitch contour is created by modulating the input contour by the randomly generated MSs. In this study, we extend this framework to ADT (i.e., NDT: \textit{neural double-tracking}). Since the GMMN-based post-filter provides natural inter-utterance variation, our NDT achieves naturally layered singing voices. The experimental evaluation demonstrates that our post-filtering approach can provide perceptible inter-utterance pitch variations while preserving speech quality and that the sound of our NDT is perceptually closer to that of natural DT than conventional signal processing-based ADT is.

\vspace{-5pt}
\section{Conventional Methods} \label{sec:conventional}
\vspace{-4pt}

\subsection{DNN-based singing voice synthesis} \label{subsec:DNNsinging}
\vspace{-4pt}
In DNN-based singing voice synthesis \cite{nishimura2016singing}, the relation between a musical score and the speech parameters of the parallel singing voice is modeled with DNNs. First, a musical score is converted into a sequence of vectors representing linguistic and musical contexts. Using DNNs, the parameters of the singing voices are predicted from the context sequence. Here, we minimize the mean square error (MSE) \cite{nishimura2016singing} between the natural and predicted speech parameters as follows:
\begin{equation}
	L_{\mathrm{MSE}}(\y, \haty) = ||\y - \haty||^2,	
\end{equation}
where $\y$ and $\haty$ are natural and predicted speech parameter sequences lasting $T$ frames, respectively. Since we are focusing on a 1-dimensional continuous fundamental frequency ($F_0$) \cite{yu11}, we define $\y$ as a scalar value sequence $[y(1), \cdots, y(t), \cdots, y(T)]^\top$, where $y(t)$ is the continuous log-scaled $F_0$ value at frame $t$ and $\top$ is the transpose. In the synthesis, $\haty=[\hat y(1), \dots, \hat y(t),\dots, \hat y(T)]^\top$ is used as the speech parameter of the synthesized singing voice. Since this synthesis process is deterministic, the resulting sound has no inter-utterance variation.

\vspace{-5pt}
\subsection{ADT}
\vspace{-4pt}

Fig.~\ref{fig:eps/DT_ADT_NDT_1024} shows the difference between DT and ADT. DT is a recording method in which multiple performances of the same phrase are mixed in order to create a layered and rich sound \cite{brice2001music,womack2014beatles}. However, singing the same phrase twice in a similar fashion may be difficult and tiresome. ADT, which is an alternative method and only requires one recording, was originally achieved by taking a vocal signal from the sync head of a multi-track, recording it to another loop of tape which was speed varied with a slow oscillation and recording it back onto the multi-track \cite{brice2001music}. More recently, signal processing-based methods have been used; the most common one uses a chorus effect \cite{izhaki2017mixing}. In the chorus effect, a waveform is copied and its pitch is modulated with a low-frequency oscillator; i.e., a sine wave is added to the original pitch contour, and the modulated sound is mixed with the original sound with some temporal difference to increase the doubled-voice feeling. Although such methods can be applied to DNN-based singing voice systems, they involve adding two signals with similar phases, which results in comb-filtering and its subsequent tonal alterations and timbre coloration \cite{izhaki2017mixing}.

\vspace{-5pt}
\section{Proposed GMMN-based post-filter and its application to NDT} \label{sec:proposed}
\vspace{-4pt}
Here, we describe the proposed GMMN-based post-filter and how NDT is performed on MSE-based singing voices. First, the MSs of the log-scaled continuous $F_0$ of natural and generated pitch contours are extracted in order to capture the temporal structure of the contours. Second, A GMMN is trained to sample the MS randomly so that naturally varied pitch contours can be produced. Finally, NDT is performed using the non-filtered and filtered voices.

	\drawfig{t}{0.75\linewidth}{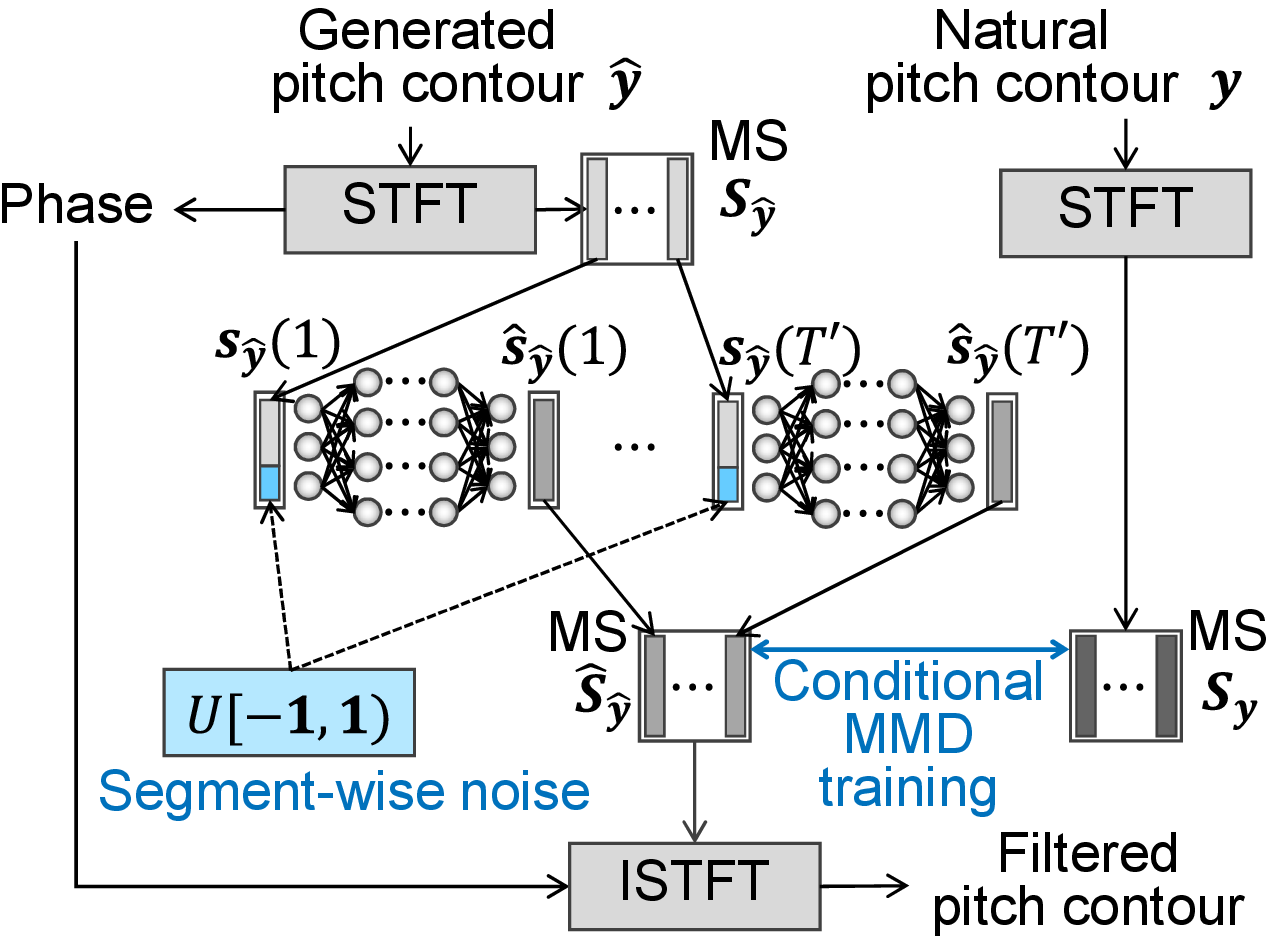} 
    {Schematic diagram of our post-filter.}

\vspace{-5pt}
\subsection{MS extraction}\label{subsec:MS}
\vspace{-4pt}
The MS is defined as the log-scaled power spectrum of a speech parameter sequence \cite{takamichi16mspf}. Namely, it captures temporal structures through the Fourier transform. We calculate the MS $\mats_\y$ of $\y$ through a short-time Fourier transform (STFT), as follows:
\begin{align}
	\Vec{\mats_y} &= [\Vec{s_y}(1), \cdots, \Vec{s_y}(\tau), \cdots, \Vec{s_y}(T')], \\
	\Vec{s_\y}(\tau) &= [s_\y (\tau , 0), \cdots, s_\y (\tau, m), \cdots, s_\y (\tau, M) ]^\top,
\end{align}
where $\tau$ is the segment index (one segment corresponds to one windowed continuous $F_0$ contour) and $m$ is the modulation frequency index. $s_\y (\tau, m)$ indicates the MS of the modulation frequency $m$ at segment $\tau$. $T'$ and $M$ correspond to the total number of segments and one-half the number of segments of the STFT, respectively. $\mats_{\haty}$, the MS of $\haty$, is calculated in the same manner. Here, the zero-mean continuous $F_0$ sequence \cite{takamichi16mspf} is used to deal with errors caused by zero padding. We reconstruct the continuous $F_0$ sequence from the inverse STFT using the MS and original phase information and use the analysis settings (e.g., windowing length) to achieve a perfect reconstruction from the STFT and inverse STFT. We use only the lower modulation frequency components (i.e., components corresponding to slowly changing temporal structures) for post-filtering, because post-filtering the higher ones causes unnatural temporal fluctuations in the $F_0$ contours.

\vspace{-5pt}
\subsection{GMMN-based post-filtering}
\vspace{-4pt}
Here, we describe the GMMN-based post-filtering method to randomly modulate the generated continuous $F_0$ sequence $\haty$. Fig.~\ref{fig:eps/postfilter_1028_2} shows the schematic diagram of the post-filter. The GMMN \cite{li15momentmatchingnetwork} is a deep generative model, and it enables stabler training compared with a GAN \cite{goodfellow14gan}. The training criterion is called maximum mean discrepancy (MMD), which is a moment-based discrepancy between two distributions. The DNN takes a prior noise vector as input; this vector is the source of the inter-utterance variation in the post-filtering stage. In the training stage of the post-filter, given the MS of the generated continuous $F_0$ and segment-wise prior noise, the conditional distribution of the MS of natural continuous $F_0$ is modeled with the DNN. Let $\Vec{n}(\tau) \sim U[-\Vec{1}, \Vec{1})$ be the prior noise vector at segment $\tau$, and $G(\cdot)$ be a DNN for post-filtering. The input of the DNN at segment $\tau$ is the joint vector $[\Vec{s}_\haty(\tau)^\top, \n(\tau)^\top]^\top$, and the output is the filtered MS $ \hat{\Vec{s}}_{\haty} (\tau)$, i.e., $ \hat{\vecs}_{\haty} (\tau)=G([\Vec{s}_\haty(\tau)^\top, \n(\tau)^\top]^\top)$. Let $\hat{\mats}_{\haty} = [\hat{\vecs}_{\haty} (1), \cdots, \hat{\vecs}_{\haty} (\tau), \cdots, \hat{\vecs}_{\haty} (T')]$; the following conditional MMD (CMMD) \cite{ren16conditionalmomentmatching} is minimized in training:
\vspace{-4pt}
\begin{align}
	L_{\mathrm{CMMD}} (\mats_{\hat \y}, \mats_\y, \hat{\mats}_{\hat \y}) 
		=& \frac{1}{T'^2} \{ \mathrm{tr}( \Vec{L}_{\mats_{\hat \y}} \cdot \Vec{K}_{\mats_\y, \mats_\y}) \nonumber \\
&+ \mathrm{tr}( \Vec{L}_{\mats_{\hat \y}} \cdot \Vec{K}_{\hat{\mats}_{\hat \y}, \hat{\mats}_{\hat \y}})  \nonumber \\
&-2\cdot \mathrm{tr}( \Vec{L}_{\mats_{\hat \y}} \cdot \Vec{K}_{\mats_\y, \hat{\mats}_{\hat \y}}) \},
\end{align}
\vspace{-11pt}
\begin{align}
\Vec{L}_{\mats_{\hat \y}} &= 
\Vec{\tilde{H}}_{ \mats_{\hat \y}}^{-1} 
\Vec{H}_{\mats_{\hat \y}} 
\Vec{\tilde{H}}_{ \mats_{\hat \y}}^{-1}, \label{lsyhat}
\end{align}
\vspace{-11pt}
\begin{align}
	\Vec{\tilde{H}}_{ \mats_{\hat \y}} &= \Vec{H}_{\mats_{\hat \y}} + \lambda \Vec{I}_{T'},
\end{align}
where $\Vec{I}_{T'}$ is the $T'$-by-$T'$ identity matrix and $\lambda$ is a regularization coefficient. $\Vec{K}_{\hat{\mats}_{\haty}, \mats_\y}$ is the $T'$-by-$T'$ Gram matrix between $\hat{\mats}_{\haty}$ and $\mats_\y$; i.e., its ${i,j}$th component is $k(\hat{\vecs}_{\haty}(i), \vecs_\y(j))$, where $k(\cdot)$ is an arbitrary kernel function between two vectors. Similarly, $\Vec{H}_{\mats_{\haty}}$ is the Gram matrix for $ \mats_{\haty}$; i.e., its ${i,j}$th component is $h(\vecs_{\haty} (i), \vecs_{\haty} (j))$, where $h(\cdot)$ is an arbitrary kernel function between two vectors. Note that we can choose different kernel functions between $k(\cdot)$ and $h(\cdot)$. After training, the model represents the natural MS distribution (i.e., variation of the pitch contours) given the generated MS. The synthesis stage first generates $\haty$ from DNN-based singing voice synthesis and calculates its MS and phase. It then filters the MS, for it to have natural variation, by using the trained GMMN and randomly sampled prior noise. The final $F_0$ contour is generated by performing the inverse STFT on the filtered MS and the non-filtered phase.

\vspace{-5pt}
\subsection{Application to NDT}
\vspace{-4pt}
Here, we describe how to give natural layeredness to a synthesized singing voice. After the speech parameter generation, one waveform is synthesized in the standard vocoding process. Another waveform is synthesized using $F_0$ values modulated by our post-filter. The final double-tracked voice is obtained by mixing the modulated waveform with the non-filtered one with some temporal difference.

\vspace{-5pt}
\subsection{Discussion}
\vspace{-4pt}

In \cite{takamichi17moment}, we built a GMMN-based spectrum generator by using frame-wise noise vectors in text-to-speech synthesis. However, the method had two problems: 1) such frame-wise noise and variation modeling caused unpleasant sounds in the case of $F_0$ contours and 2) the GMMN conditioned with linguistic features could not provide perceptible variations because of the sparseness of the linguistic features. This paper's method effectively solves these problems. This is because the MS of $F_0$ contours is a lower-dimensional and effective representation with which to capture segment-wise temporal structure, and filtering the lower modulation frequency components can modulate the $F_0$ contour and at the same time preserve the continuity of the contour. The other reason is that the use of GMMNs as a post-filter can avoid the sparseness problem and can provide perceptible inter-utterance variations, as shown in Section \ref{subsec:Dif}.

Since our method considers the distribution of natural MSs, the variation of the post-filtered voices should be in the natural range. Fig.~\ref{fig:eps/4lines_1024_2} shows MSE-based (i.e., deterministic and non-filtered) and post-filtered (i.e., randomly modulated) pitch contours. We can see that our method samples random, but continuous pitch contours.
This is the first study to 1) provide natural inter-utterance pitch variations by using GMMNs to model MSs and to 2) introduce such a system as a post-filter for singing voices.
As described in Section 1, inter-utterance variation enables one to pick a favorite from various voices. Since the pitch contour of each segment can be saved by fixing the input noise, users can choose their favorite contour phrase-by-phrase and concatenate phrases to make the whole song.

Conventional ADT involves a waveform made by modulating the original sound deterministically without considering the natural distribution of pitch variations. NDT is a new approach in that the sound is modulated considering the natural pitch variations and that it results in more doubled-voice feeling as shown in Section \ref{subsec:eva_NDT}.

Data augmentation is a powerful method for improving the accuracy of DNN modeling. In this study, we utilized data augmentation in relation to the STFT. Since the MS utilizes an STFT, the value changes significantly with the segment position (i.e., the frame index of the beginning of the segmentation). To cover such perturbations, we added all possible offsets to the first frame of the STFT analysis. This method effectively augments the training data of our post-filter and improves its training accuracy. 

Since our method utilizes pitch contours as the input and the output of the neural network, we expect that it can be extended to NDT for human singers, i.e., where a human singer sings only once and the extended NDT synthesizes a naturally layered voice. We will pursue this idea in the future (see Section 5).

\drawfig{t}{0.95\linewidth}{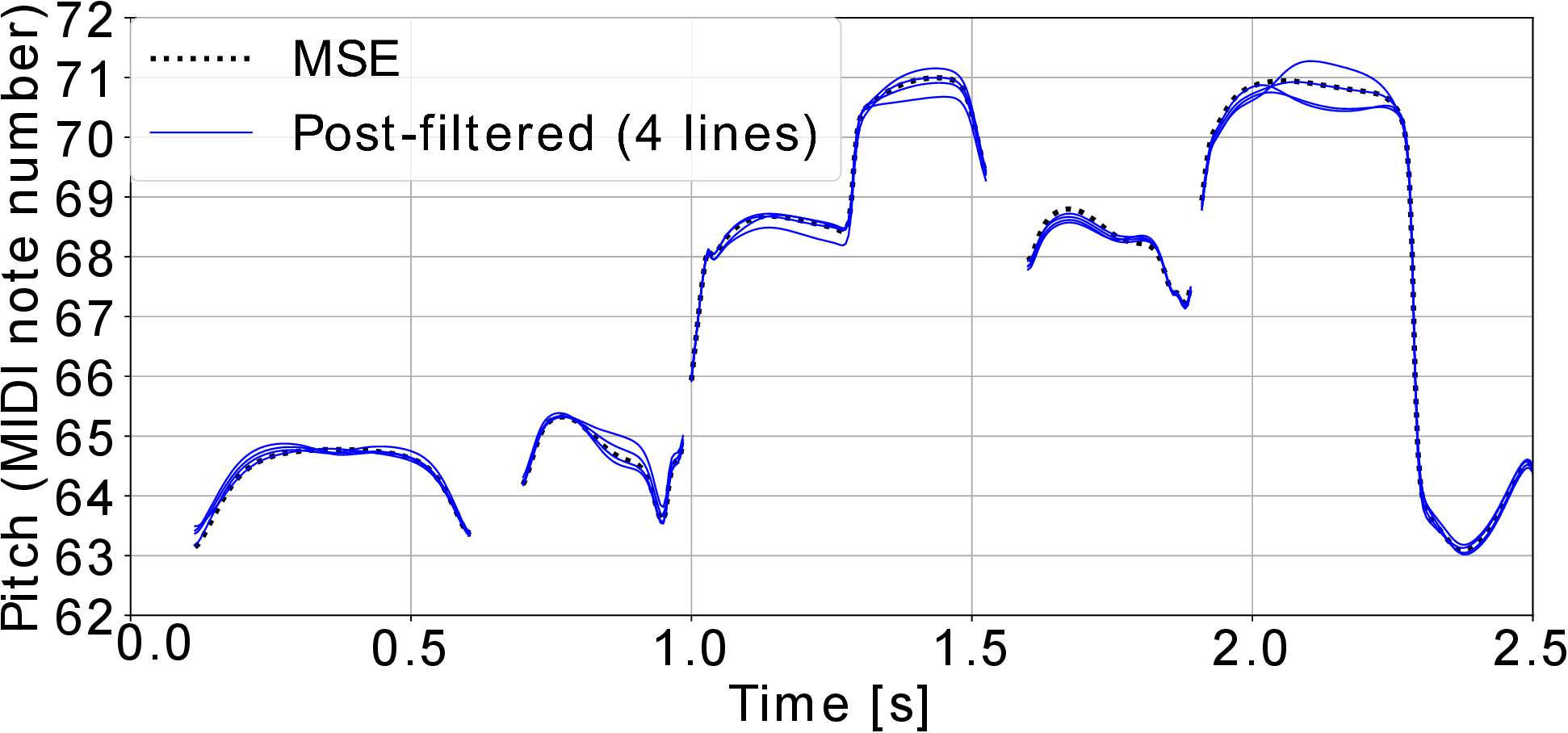} {Example of generated pitch contours. We used proposed method to sample four contours. Value of unity on vertical axis is equal to semitone.}

\vspace{-5pt}
\section{Experimental Evaluation} \label{sec:experiment}
\vspace{-6pt}
\subsection{Experimental conditions}
\vspace{-4pt}
We used Japanese singing voice data of 31 songs from a singing-voice-synthesis demo of HTS \cite{hts}, 26 songs from the JSUT-song corpus \cite{jsutsong}, and nine in-house songs sung by the same singer as in the JSUT-song corpus. From these corpora, we used 58 songs for training the DNN for singing voice synthesis, 28 songs of the HTS demo for training our post-filter, and three songs of the HTS demo that were not included in the training data for the evaluation. Label data were augmented 3-fold by transposing the songs up and down a semitone \cite{blaauw2017neural}.
The speech signals were sampled at a rate of 16 kHz. The WORLD analysis-synthesis system \cite{morise16world} was used to extract the speech features and to synthesize the waveforms. The frame shift was 5 ms. The DNN used to predict the MSE-based speech parameters was a feed-forward network including a 705-dimensional input layer, $3\times256$-unit gated linear unit (GLU) \cite{dauphin16gatedlinear} hidden layers, and a 127-unit linear output layer. The learning rate was 0.005, and the batch size was 500. The learning process was repeated for 50 epochs by using AdaGrad \cite{duchi11adagrad}. The 705-dimensional input features included 688-dimensional linguistic and musical features, a one-hot song code and a one-hot singer code \cite{hojo16speakercode}. The DNN predicted a 127-dimensional vector consisting of 40-dimensional mel-cepstral coefficients, log-scaled continuous $F_0$, band-aperiodicity \cite{kawahara01,ohtani06}, dynamic (delta- and delta-delta-) features \cite{tokuda00} of those 42-dimensional parameters, and a binary unvoiced/voiced label. The DNN used in the conditional GMMN was a feed-forward network including an 11-dimensional input layer, $3\times128$-unit GLU hidden layers, and input-to-output residual net \cite{kaneko2017generative}. The learning process was repeated ten times by using AdaGrad. The learning rate was 0.005, and the batch size was 13000.
We used an approximation using 1024-dimensional random Fourier features \cite{rahimi2008random} to calculate $\Vec{L}_{\mats_{\hat \y}}$ in Eq. (\ref{lsyhat}) because calculating the inverse matrix was computationally infeasible.
The 11-dimensional input vector consisted of the first-order MS ($m=1$) of the singing voice synthesized based on MSE and a ten-dimensional noise vector generated from a uniform distribution $U[-\textbf{1}, \textbf{1})$. For stable training, noise vectors were first generated for each segment and were then fixed during training. The regularization coefficient $\lambda$ was set to 0.01. The kernel function was Gaussian, i.e., $\mathrm{exp}\{-||\vecs_\y(i)-\hat{\vecs}_\haty(j)||^2/\sigma^2\}$. A Gaussian kernel was used for the input features as well. The $\sigma$ for the input features was 100.0, and the $\sigma$ for the output features was 1.0; these values were empirically chosen. The natural MS was normalized so that all data fell within the range $[0.01, 0.99]$. For the STFT, a 96-frame (480 ms) Hanning window and 48-frame (240 ms) segment shift were used. To clarify the effect of pitch modulation, we used spectral parameters, band-aperiodicity, and the unvoiced/voiced label of the natural singing voices in the vocoding process.

We conducted three subjective evaluations for determining 1) whether our post-filter provided perceptible inter-utterance variations, 2) whether the post-filter degraded the naturalness of the synthesized voices, and 3) whether the sound of NDT was perceptually closer to that of natural DT than ADT was. The evaluations were performed using the Lancers crowdsourcing platform \cite{lancers}. To make it easier for listeners to judge, we manually split the songs into segments in correspondence with three conditions: short (one phrase), middle, and long (several phrases). The average lengths for the three conditions were 3.01 s, 4.88 s, and 10.24 s, respectively. 

\begin{table}[tb]
\centering
\caption{Answer rate of perceived inter-utterance difference}
\vspace{-3mm}
{\renewcommand\arraystretch{1.2}
 \begin{tabular}{ccc} \\ \hline \hline
 Proposed & MSE & \textit{p}-value \\ \hline
 \textbf{0.276} & 0.176 & 7.45$\times10^{-3}$ \\ \hline \hline
 \end{tabular}
 }
 \label{tab:Dif}
\vspace{-6mm}
\end{table}

\vspace{-11pt}
\subsection{Perception of inter-utterance pitch variation}
\label{subsec:Dif}
\vspace{-4pt}
To determine whether the inter-utterance pitch variation could be perceived by human listeners, we asked 25 listeners whether they felt there was a difference between a pair of singing voices. The short-duration voices were used because it is easier to remember subtle differences if shorter sounds are presented. Each participant listened to 20 pairs of singing voices consisting of ten pairs of randomly post-filtered voices (proposed) and ten pairs of identical MSE-based voices. We used Welch's \textit{t} test to calculate the \textit{p}-value.

Table \ref{tab:Dif} shows the results. The perception rate of MSE was 17.6\% despite that the same voices were presented. On the other hand, the rate of our method was statistically significantly higher than that of MSE. This suggests that our method is capable of producing perceptible inter-utterance variations.

\vspace{-8pt}
\subsection{Naturalness of post-filtered voice}
\label{subsec:Natural}
\vspace{-4pt}
We tried to determine whether our post-filter degraded the quality of the synthesized voices. We asked 25 participants to listen to ten pairs of post-filtered and MSE-based voices and to choose the more natural one. The middle- and long-duration voices were used to make it easy to judge the overall naturalness.

Table \ref{tab:Nat} shows the results. There was no statistically significant difference for either the middle-duration voices or the long ones. This implies that the post-filter did not degrade the naturalness of the singing voices.

\begin{table}[tb]
\caption{Preference scores of singing voice naturalness and their \textit{p}-values for the middle and long conditions}
\centering
{\renewcommand\arraystretch{1.2}
\begin{tabular}{c|ccc}
\hline
\hline
 Length condition & Proposed & MSE & \textit{p}-value \tabularnewline \hline
 Middle & 0.504 & 0.496 & $8.58\times 10^{-1}$ \tabularnewline
 Long & 0.480 & 0.520 & $3.72\times 10^{-1}$ \tabularnewline
 \hline
 \hline
 \end{tabular}
 }
\label{tab:Nat}
\vspace{-6mm}
\end{table}

\begin{table}[tb]
\centering
\caption{Preference scores of double-trackedness and their \textit{p}-values for the middle and long conditions}
{\renewcommand\arraystretch{1.2}
 \begin{tabular}{c|ccc} \hline \hline
 Length condition & NDT & ADT & \textit{p}-value \\ \hline
 Middle & \textbf{0.724} & 0.276 & $<10^{-10}$ \\ 
 Long & \textbf{0.736} & 0.264 & $<10^{-10}$ \\ \hline \hline
 \end{tabular}
 }
 \label{tab:DT}
\vspace{-6mm}
\end{table}

\vspace{-5pt}
\subsection{Evaluation of NDT}
\label{subsec:eva_NDT}
\vspace{-4pt}
We evaluated the \textit{double-trackedness} (i.e., the perceptual similarity to naturally double-tracked sound) of the proposed NDT and conventional ADT:
\vspace{-4pt}
\begin{description}
\item[NDT:] Modulating the log-scaled $F_0$ sequence using our post-filter and mixing its vocoded speech waveform with the MSE-based waveform.
\vspace{-4pt}
\item[ADT:] Modulating the log-scaled $F_0$ sequence by using a low-frequency oscillator and mixing its vocoded speech waveform with the MSE-based waveform. The shape, rate, and depth of the oscillation were a sine wave, 0.775 Hz, and 10\% of a semitone, respectively. The parameters were determined by referring to \cite{izhaki2017mixing}. The modulation was performed in the vocoder parameter domain, not in the waveform domain, because doing so is considered to produce fewer artifacts.
\end{description}
\vspace{-4pt}
In both methods, the modulated wave was delayed by 20 ms and the volume was reduced by 3 dB to produce the usual ADT setting \cite{izhaki2017mixing}. We asked 25 participants to listen to ten pairs of sounds generated under the two conditions above and to choose the one that sounded more like a naturally double-tracked sound. As before, the middle- and long-duration voices were used.

Table \ref{tab:DT} shows the results. The score of our NDT was significantly higher than that of conventional ADT for both the middle and long conditions. This suggests that our deep-generative method gives a more double-tracked feeling than the conventional signal processing-based approach does.

\vspace{-6pt}
\section{Conclusion} 
\label{sec:conclusion}
\vspace{-4pt}
This paper described a GMMN-based post-filter for generating inter-utterance pitch variation and its application to ADT. Although a human singer never sings the same song in the same way twice, conventional singing voice synthesis systems only generate a single waveform that is considered to be the most natural one. Our post-filter models inter-utterance pitch variation by using a GMMN to match the statistical moments of the MS of synthesized voices and those of natural voices. Experimental results suggest that the post-filter can generate pitch variations that are perceptible by human listeners without degrading the naturalness of the synthesized singing voices. We also discussed the effectiveness of applying the post-filter to ADT. Experimental results suggest that the proposed NDT is perceptually closer to natural DT than conventional ADT is.

Future work will include modeling the inter-utterance variation of the duration and spectral parameters and combining them with that of pitch in order to reproduce more natural variation. Another topic is formulating a post-filter that can be used to modulate the MS of natural singing voices.

\textbf{Acknowledgements:}
Part of this work was supported by SECOM Science and Technology Foundation.

\vfill\pagebreak

\bibliographystyle{IEEEbib}
\bibliography{Template}

\end{document}